\documentclass{webofc}
\usepackage[varg]{txfonts}   % Web of Conferences font
\newcommand{\be}{\begin{equation}}
\newcommand{\ee}{\end{equation}}
\newcommand{\bea}{\begin{eqnarray}}
\newcommand{\eea}{\end{eqnarray}}

\newcommand{\ttc}[1]{\multicolumn{2}{c}{#1}}

\begin{document}
\title{Hadronic light-by-light scattering in the muon $g-2$}

\author{\firstname{Andreas} \lastname{Nyffeler}\inst{1}\fnsep\thanks{\email{nyffeler@uni-mainz.de}}}

\institute{PRISMA Cluster of Excellence and Institut f\"ur Kernphysik,
  Johannes Gutenberg-Universit\"at, \\ 
  D-55128~Mainz, Germany}

\abstract{%
  We briefly review the current status of the hadronic light-by-light
  scattering contribution to the anomalous magnetic moment of the
  muon. Based on various model calculations in the literature, we
  obtain the estimate $a_\mu^{\mathrm{HLbL}} = (102 \pm 39) \times
  10^{-11}$. Recent developments including more model-independent
  approaches using dispersion relations and lattice QCD, that could
  lead to a more reliable estimate, are also discussed.}

%%%%%%%%%%%%%%%%%%%%%%%%%%%%%%%%%%%%%%%%%%%%%%%%%%%%%%%%%%%%%%%%%%%%%%%%%%%%
% extra for preprint
\thispagestyle{empty}
\begin{flushright}
{\large
October 2017 \\[1mm] 
MITP/17-074}
\end{flushright}

\vfill

\begin{center}
{\LARGE\bf
Hadronic light-by-light scattering in the muon $g-2$}\\[1.5cm]  
{\large Andreas Nyf\/feler$^{\dagger, \star}$}\\[1cm]
{\large PRISMA Cluster of Excellence and Institut f\"ur Kernphysik, \\[1mm] 
Johannes Gutenberg-Universit\"at, D-55128 Mainz, Germany}

\vfill

{\large\bf Abstract}\\[3mm]

\begin{minipage}{0.825\textwidth}
  We briefly review the current status of the hadronic light-by-light
  scattering contribution to the anomalous magnetic moment of the
  muon. Based on various model calculations in the literature, we
  obtain the estimate $a_\mu^{\mathrm{HLbL}} = (102 \pm 39) \times
  10^{-11}$. Recent developments including more model-independent
  approaches using dispersion relations and lattice QCD, that could
  lead to a more reliable estimate, are also discussed.
\end{minipage}
\end{center}

\vfill
\noindent\rule{8cm}{0.5pt}\\
$^\dagger$Invited talk at International Workshop on $e^{+} e^{-}$
collisions from Phi to Psi 2017, 26-29 June 2017, Mainz, Germany. \\
$^\star$e-mail: nyffeler@uni-mainz.de 
\setcounter{page}{0}
\newpage
%
% end extra for preprint
%%%%%%%%%%%%%%%%%%%%%%%%%%%%%%%%%%%%%%%%%%%%%%%%%%%%%%%%%%%%%%%%%%%%%%%%%%% 

\maketitle

\section{Introduction}
\label{sec:Intro}

Since 1947, the anomalous magnetic moments of the electron and the
muon have always triggered new approaches and developments in loop
calculations in quantum field theories~\cite{JN_09,
  talk_Steinhauser}. The muon $g-2$ thereby serves as an important
precision test of the Standard Model (SM)~\cite{JN_09,
  talks_Teubner_Zhang_Jegerlehner}.  For some time already, there is a
discrepancy $a_\mu^{\mathrm{exp}} - a_\mu^{\mathrm{SM}} \approx (300
\pm 80) \times 10^{-11}$ of $3-4\sigma$ between experiment and
theory. This could be a sign of New Physics, but the theoretical
uncertainties from hadronic vacuum polarization (HVP) and hadronic
light-by-light scattering (HLbL) need to be better controlled to draw
firm conclusions. This issue is even more pressing in view of more
precise new experiments~\cite{talks_Lee-Roberts_Mibe} at Fermilab and
J-PARC, with a precision goal of $\delta a_\mu^{\mathrm{exp}} = 16
\times 10^{-11}$ in a few years time.

For the HLbL contribution, the following estimates are frequently
used:   
\bea
a_\mu^{\mathrm{HLbL}} & = & (105 \pm 26) \times 10^{-11}, \qquad
\mbox{\cite{PdeRV_09}}, \quad \mbox{(``Glasgow
  consensus'')}, \label{HLbL_PdeRV} \\   
a_\mu^{\mathrm{HLbL}} & = & (116 \pm 39) \times 10^{-11}, \qquad
\mbox{\cite{N_09,JN_09}}.   \label{HLbL_JN} 
\eea
Note that they are both based on almost the same input from
calculations by various groups using different hadronic
models~\cite{HKS, BPP, KN_02, MV_04}, which suffer from uncontrollable
uncertainties. More model-independent approaches, using dispersion
relations (data driven)~\cite{HLbL_DR_Bern,
  HLbL_DR_Bern_recent, HLbL_DR_Mainz, HLbL_sum_rules,
  Danilkin_Vanderhaeghen_17} and lattice
QCD~\cite{Lattice_HLbL_Blum_et_al, Lattice_HLbL_Mainz,
  HLbL_sum_rules_lattice}, have been proposed and first promising
results have been obtained recently. To come up with a more refined
estimate for HLbL with a controlled uncertainty is also one of the
goals of a recently formed ``Muon $g-2$ Theory
Initiative''~\cite{talk_El-Khadra} that will accompany the upcoming
new experiments.

\section{Current status of HLbL: model calculations}
\label{sec:HLbL_models}

\begin{figure}[t]
\includegraphics[width=\textwidth]{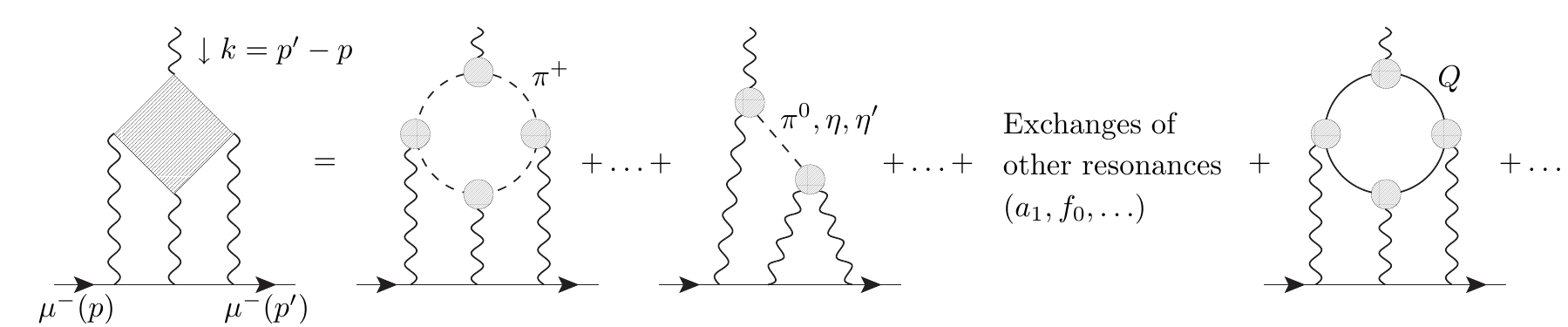}    

\vspace*{0.2cm}
{\small 
\hspace*{0.32cm}Contribution: \hspace*{1.5cm} pion-loop  \hspace*{1.15cm}
pseudoscalar  \hspace*{3.8cm}  quark-loop \\{}  
\hspace*{3.65cm} (dressed) \hspace*{1.35cm} exchanges \hspace*{4.05cm}
(dressed) \\{} 
\hspace*{0.32cm}Chiral counting: \hspace*{1.6cm}  $p^4$  \hspace*{2.3cm}
$p^6$  \hspace*{2.5cm} $p^8$  \hspace*{2.1cm} $p^8$ \\{}  
\hspace*{0.32cm}$N_c$ counting:   \hspace*{2.1cm}  $1$   \hspace*{2.4cm}
$N_c$  \hspace*{2.45cm} $N_c$  \hspace*{2.1cm} $N_c$  
}
\caption{The different contributions to HLbL scattering in the muon
  $g-2$ and their chiral and large-$N_c$ counting.}  
\label{fig:HLbL}
\end{figure}

The HLbL contribution to the muon $g-2$ contains the QCD correlation
function of four hadronic electromagnetic currents, connected by
off-shell photons to the muon line, see Figure~\ref{fig:HLbL}. Within
a hadronic picture, the four-point function is decomposed into
single-meson exchanges and loops of hadrons, e.g.\ pions. Often the
QCD short-distance part is modelled by a dressed constituent quark
loop, which raises issues of double counting. The couplings of the
hadrons (and the constituent quarks) to the photons involve, in
general, momentum dependent vertex functions (form factors). The
different contributions have been classified in
Ref.~\cite{deRafael_94} according to their leading order in the chiral
expansion $p^2$ and their large-$N_c$ counting to bring some order and
systematics into the calculations.

The relevant momentum scales in HLbL are around $0-2~\mbox{GeV}$,
i.e.\ in the non-perturbative resonance region of QCD. The QCD
four-point function is, however, a very complicated object that
involves many Lorentz structures~\cite{BPP, HLbL_DR_Bern} that depend
on several invariant photon momenta with mixed regions of small and
large momenta. Therefore the distinction between low and high energies
and the use of an effective field theory approach (chiral perturbation
theory with hadronic resonances) at low momenta and of perturbative
QCD at high momenta is not so straightforward.  So far only hadronic
models have been used to estimate the full HLbL contribution. A
selection of these results and some compilations, including those
quoted in Eqs.~(\ref{HLbL_PdeRV}) and (\ref{HLbL_JN}), is shown in
Table~\ref{tab:HLbL_models}. One important difference between these
two compilations is the combination of the errors of the individual
contributions. They are combined in quadrature in Ref.~\cite{PdeRV_09}
and linearly in Refs.~\cite{N_09, JN_09}, as was done in
Ref.~\cite{BPP}. Since these are model errors, not experimental
uncertainties, both ways of combining them can be questioned.

\begin{table}[h]
\centering
\caption{Selection of model estimates for the various contributions to
  $a_\mu^{\mathrm{HLbL}} \times 10^{11}$.}   
\label{tab:HLbL_models}
\renewcommand{\arraystretch}{1.5}
\begin{tabular}{cr@{$\pm$}lr@{$\pm$}lr@{$\pm$}lr@{$\pm$}lr@{$\pm$}lr@{$\pm$}lr@{$\pm$}l}
\hline%\noalign{\smallskip}
Contribution & \ttc{\cite{HKS}} & \ttc{\cite{BPP}} & 
\ttc{\cite{KN_02}} & \ttc{\cite{MV_04}} & \ttc{\cite{HLbL_2007}} &
\ttc{\cite{PdeRV_09}} & \ttc{\cite{N_09, JN_09}} \\  
\hline%\noalign{\smallskip}
 $\pi^0,\eta,\eta'$ & $82.7 $&$ 6.4$ & $85$ & $13$ & $ 83 $&$ 12$
& $114 $&$ 10$ & \ttc{$-$} & $114$ & $13$ & 99 & 16 \\ 
axial vectors & $1.7 $&$ 1.7$ & $2.5 $&$ 1.0$ &\ttc{$-$} & 
$22 $&$ 5$& \ttc{$-$} & 15 & 10 & 22 & 5 \\ 
scalars       & \ttc{$-$} & $-6.8$&$ 2.0$ & \ttc{$-$} & \ttc{$-$}&
\ttc{$-$} &  $-7$ & $7$ & $-7$&$2$ \\ 
$\pi,K$ loops &  $-4.5$&$ 8.1$ & $-19$ &$ 13$ &\ttc{$-$} & \ttc{$-$} &
 \ttc{$-$} & $-19$ & $19$ &$-19 $&$ 13$ \\ 
$\pi,K~\mbox{loops} \atop + \mbox{subl.}~N_C$ & \ttc{$-$} & \ttc{$-$} &
\ttc{$-$} & $0$&$ 10$ & \ttc{$-$} & \ttc{$-$} & \ttc{$-$} \\ 
quark loops  & $9.7 $&$ 11.1$ & $21 $&$ 3$ & \ttc{$-$} & \ttc{$-$} & \ttc{$-$}
&  \ttc{$2.3$ (c-quark)} & $21 $&$ 3$ \\
\hline 
Total & $89.6 $&$ 15.4$ & $83$&$ 32$ & $80 $& $40$ &
$136$ & $25$  & $110$ & $40$ & 105 & 26 & 116 & 39 \\ 
%
%%\noalign{\smallskip}
\hline
\end{tabular}
\end{table}

The contribution from the light pseudoscalars $\pi^0,\eta,\eta^\prime$
is numerically dominant according to most model calculations. Because
of this observation, there are many evaluations of this contribution,
see Refs.~\cite{JN_09, Bijnens_15, N_16} and references therein. The
central value is about $a_{\mu}^{\mathrm{HLbL; P}} = 90 \times
10^{-11}$ with a spread for most calculations of about 15\% (but 30\%
if the central values of all estimates are taken into account), which
can be understood by looking at the relevant momentum regions in a
3-dimensional integral representation~\cite{JN_09, N_16}, with
model-independent weight functions that are peaked below
$1~\mbox{GeV}$ for the pion and below about $1.5 - 2~\mbox{GeV}$ for
$\eta, \eta^\prime$. As long as the transition form factors fall off
for large momenta, one obtains always very similar results. In
Ref.~\cite{MV_04} a QCD short-distance constraint from the operator
product expansion (OPE) on the four-point function was derived by
connecting it to the chiral triangle anomaly. The constraint is then
saturated by the pion-pole contribution alone which is a model
assumption. This leads to an increased value, since there is no pion
transition form factor at the external vertex, but then no quark-loop
contribution should be added.

The other contributions to HLbL are, however, not negligible at the
level of the precision goal of $(15-20) \times 10^{-11}$ needed to
match future experiments. For the dressed pion-loop there is a strong
model-dependence, cf.\ the results obtained in Refs.~\cite{HKS, BPP}
as discussed in Refs.~\cite{MV_04, BZ-A}. There is also some numerical
cancellation with the dressed quark-loop. On the other hand, recent
reevaluations~\cite{Pauk_Vanderhaeghen_mesons_14, Jegerlehner_14,
  Jegerlehner_15} of the axial-vector contribution lead to a much
smaller estimate $a_{\mu}^{\mathrm{HLbL; axial}} = (8 \pm 3) \times
10^{-11}$ than in Ref.~\cite{MV_04}. Using these new evaluations and
the observation that the contribution from tensor mesons seems to be
very small, $a_\mu^{\mathrm{HLbL; tensor}} = 1 \times
10^{-11}$~\cite{Pauk_Vanderhaeghen_mesons_14,
  Danilkin_Vanderhaeghen_17}, lead us to suggest the following update
of our earlier estimate~\cite{N_09, JN_09} for the HLbL contribution
(see also Ref.~\cite{Jegerlehner_15}):
\be \label{HLbL_update}
a_{\mu}^{\mathrm{HLbL}} = (102 \pm 39) \times 10^{-11}. 
\ee
We have only updated the central value and kept the error estimate
unchanged. Using the new estimate for the axial-vectors would also
shift the ``Glasgow consensus'' downward to $a_{\mu}^{\mathrm{HLbL}} =
(98 \pm 26) \times 10^{-11}$.

\section{Model-independent approaches to HLbL}
\label{sec:HLbL_model-independent} 

\subsection{HLbL from dispersion relations (data driven)}

The approach with dispersion relations (DR) for HLbL proposed in
Refs.~\cite{HLbL_DR_Bern, HLbL_DR_Bern_recent,
  HLbL_DR_Mainz, HLbL_sum_rules, Danilkin_Vanderhaeghen_17} tries to
relate parts of the contributions in the Feynman diagrams in
Figure~\ref{fig:HLbL} from on-shell intermediate states, e.g.\ from
the pseudoscalar-poles and from two pions (pion-loop), to in principle
measurable form factors and cross-sections $\gamma^*\gamma^* \to
\pi^0,\eta,\eta^\prime$ and $\gamma^*\gamma^* \to \pi^+\pi^-,
\pi^0\pi^0$. These contributions involving the lightest hadrons are
expected to dominate numerically based on experiences of other uses of
DR's. It is also confirmed by the results from the model calculations
in Table~\ref{tab:HLbL_models}. Note that in a quantum field
theoretical approach, the Feynman diagrams in Figure~\ref{fig:HLbL}
contain off-shell hadrons and the individual contributions to HLbL are
model-dependent~\cite{JN_09}. This complicates the comparison of the
results for the individual contributions in different models.  The
dispersive approach in Refs.~\cite{HLbL_DR_Bern, HLbL_DR_Bern_recent}
considers first the fully off-shell four-point function and projects
on the on-shell intermediate hadronic states with one or two
pions. Only at the end the contribution to the muon $g-2$ is
evaluated.  On the other hand, the approach in
Ref.~\cite{HLbL_DR_Mainz} starts from a DR for the Pauli form factor
$F_2(k^2)$ and then evaluates $a_\mu^{\mathrm{HLbL}} = F_2(k^2 = 0)$.

Assuming that experimental results for the two-photon processes above
can be obtained directly, which is not yet the case at present for two
off-shell photons, or indirectly, using other DR's with purely
hadronic processes or single-virtual photons, the hopefully
numerically dominant contributions to HLbL from single light
pseudoscalars and from the two-pion intermediate states, can be
obtained with high precision, e.g.\ better that 10\%, where the error
largely relies on experimental input only, like for the HVP.  On the
other hand, it should be possible to obtain the contributions from the
presumably numerically subdominant $3\pi$-intermediate states, e.g.\
axial-vectors, from further multi-pion states (heavier resonances) and
from the dressed quark-loop from models and theoretical constraints,
e.g.\ by matching with perturbative QCD, to 30\% to obtain an overall
error of about 20\%~($\delta a_\mu^{\mathrm{HLbL}} \approx 20 \times
10^{-11}$ if the central value stays the same as in
Eq.~(\ref{HLbL_update})).

For the pseudoscalar-pole contribution to HLbL the double-virtual
transition form factors are needed as input, see
Figure~\ref{fig:HLbL}. There exist already several measurements of the
single-virtual transition form factors in certain momentum
regions~\cite{TFF_single_virtual}. The double-virtual form factors
have only been modelled so far. They can hopefully be measured at
BESIII~\cite{N_16} or they can be obtained from a DR
itself~\cite{TFF_DR} or from lattice QCD~\cite{TFF_lattice}. It
remains to be seen, which precision on the pseudoscalar-pole
contribution can be obtained in this way~\cite{N_16, TFF_lattice} and
how much modelling will still be needed, e.g. for the high-energy
region in the DR's or to parametrize experimental or lattice data.

Very recently, a first estimate for the $2\pi$-contribution has been
obtained in Ref.~\cite{HLbL_DR_Bern_recent}. The contribution itself
is split into two parts. The first is the pion-box contribution, see
the first diagram on the right-hand side of Figure~\ref{fig:HLbL},
which has a one-pion cut in the $s$- and the $t$-channel and is
identical to scalar QED with all vertices dressed by the pion vector
form factor obtained from a DR itself. The second part describes the
remaining $\pi\pi$-rescattering effects. In a first approximation only
$S$-wave rescattering from the pion-pole in the left-hand cut (lhc)
have been taken into account and the high-energy part in the DR is
modelled. The results read:
\bea
 a_\mu^{\rm \pi-{\rm box}} & = & -15.9(2) \times
 10^{-11},  \label{pion_box} \\   
 a_{\mu,J=0}^{\rm \pi\pi,\pi-{\rm pole~lhc}} & = & -8(1) \times
 10^{-11},  \\
a_\mu^{\rm \pi-{\rm box}} + a_{\mu, J=0}^{\rm \pi\pi,\pi-{\rm
    pole~lhc}} & = & -24(1) \times 10^{-11}.  \label{pipi_DR_approx}
\eea 
The result in Eq.~(\ref{pipi_DR_approx}) is much more precise than the
values for the pion-loop given in Table~\ref{tab:HLbL_models}.  Not
surprisingly, the the pion-box contribution in Eq.~(\ref{pion_box}) is
close to the estimate from Ref.~\cite{BPP} that uses full VMD. That
evaluation was recently reanalyzed~\cite{Bijnens_15,
  Bijnens_Relefors_16} yielding
$a_\mu^{\mathrm{HLbL};\pi-\mathrm{loop}} = (-20 \pm 5) \times
10^{-11}$ close to the original value and compatible with the estimate
in Eq.~(\ref{pipi_DR_approx}). The question is, how much is not yet
included in the truncated dispersive approach compared to the complete
two-pion contribution.
  
Another approach was developed by the Mainz group by using HLbL
forward scattering sum rules~\cite{HLbL_sum_rules} to constrain, under
the assumption of factorization, the transition form factors of
various mesons with two off-shell photons from experimental data or
from lattice QCD~\cite{HLbL_sum_rules_lattice}. These transition form
factors are then used to evaluate the corresponding contributions to
HLbL in the $g-2$~\cite{Danilkin_Vanderhaeghen_17}.

\subsection{HLbL from lattice QCD}

The calculation of HLbL in lattice QCD was proposed about 10 years
ago, but only recently some first, still incomplete, numerical results
have been obtained~\cite{Lattice_HLbL_Blum_et_al}. After several
changes in the strategy, the calculation is now performed in position
space, one obtains directly $a_\mu^{\mathrm{HLbL}} = F_2(k^2 = 0)$ and
exact expressions for all photon propagators are used. The latest
result with physical pion mass, a finite lattice spacing $a^{-1} =
1.73~\mbox{GeV}$ and a box-size with $L = 5.5~\rm{fm}$ reads:
\bea
  a_\mu^\text{cHLbL} &=& (116.0 \pm 9.6) \times 10^{- 11},  \qquad 
  \ \mbox{(quark-connected diagrams),}  \\
  a_\mu^\text{dHLbL} &=& (- 62.5 \pm 8.0) \times 10^{- 11}, \qquad
  \mbox{(leading quark-disconnected diagrams).} 
\eea
The size of these estimates is in the ballpark of the model
calculations, see Table~\ref{tab:HLbL_models}. But note that the error
is statistical only. Missing systematic uncertainties are potentially
large power-law finite-volume effects from QED in a box~$\sim 1/L^2$
(this has now been overcome by using infinite volume, continuum QED in
the last paper of Ref.~\cite{Lattice_HLbL_Blum_et_al}, as proposed in
Ref.~\cite{Lattice_HLbL_Mainz}) and from the finite lattice
spacing. Also subleading quark-disconnected diagrams could be around
10\% of the numbers given. Finally, it was found empirically that the
short-distance contribution $< 0.6~\rm{fm}$ dominates the HLbL
integral.

Independently, the lattice group at Mainz~\cite{Lattice_HLbL_Mainz}
developed in the last few years an approach in position space. We
obtain the master formula
\bea 
a_\mu^{\rm HLbL} & = & \frac{m e^6}{3}  \int d^4y
      \int d^4x
      \underbrace{\bar{\cal L}_{[\rho,\sigma];\mu\nu\lambda}(x,y)}_{\rm QED}\;
      \underbrace{i\widehat\Pi_{\rho;\mu\nu\lambda\sigma}(x,y)}_{\rm
        QCD}, 
\\ 
i\widehat \Pi_{\rho;\mu\nu\lambda\sigma}( x, y) & = &  
-\int d^4z\; z_\rho\, \Big\langle \,j_\mu(x)\,j_\nu(y)\,j_\sigma(z)\,
j_\lambda(0)\Big\rangle, 
\eea 
where the QED part is computed semi-analytically in the continuum and
in infinite volume. Therefore there are no power-law $1/L^2$
finite-volume effects. We have kept Lorentz invariance manifest which
allows us to parametrize the QED kernel by six weight functions (and
derivatives thereof) that depend only on $x^2, y^2$ and $x \cdot y$
which we have pre-computed on a 3-dimensional grid. The QCD part will
be computed on the lattice eventually. 

As a numerical test on our approach we have calculated the presumably
numerically dominant pion-pole contribution to HLbL with a simple VMD
model in position space~\cite{Lattice_HLbL_Mainz}. In contrast to the
observations in Ref.~\cite{Lattice_HLbL_Blum_et_al} we find that one
needs rather large lattices of $L \sim (5-10)~\rm{fm}$ to reproduce
the known results for the physical pion mass. As a further check, we
also reproduce the known results for a lepton-loop in QED with
$m_{\mathrm{loop}} = m_\mu, 2 m_\mu$ at the percent level.

These numerical tests give us confidence in our approach and the
lattice simulations to calculate HLbL in the muon $g-2$ with full QCD
will soon begin. Note that as another complementary way to tackle HLbL
scattering, lattice QCD calculations have already been performed to
constrain hadronic models for transition form factors using HLbL
forward scattering sum rules~\cite{HLbL_sum_rules,
  HLbL_sum_rules_lattice} and to evaluate the pion transition form
factor with two off-shell photons on the lattice~\cite{TFF_lattice}.

\section{Conclusions}
\label{sec:Conclusions}

We have briefly reviewed the current approach to HLbL using hadronic
models and given an updated value $a_{\mu}^{\mathrm{HLbL}} = (102 \pm
39) \times 10^{-11}$ in Eq.~(\ref{HLbL_update}) where the uncertainty
is rather arbitrary. Hopefully, the data driven dispersive approaches
and lattice QCD will soon be able to give a reliable estimate. The
experimental results presented at this PHIPSI 2017 meeting (and to
come in the near future) from various measurements below and above the
$\phi$-meson thereby serve as important input and constraints on the
theoretical approaches. Note that in order to fill the gap between
experiment and theory by HLbL alone, the HLbL contribution would have
to be four times bigger than the above estimate, i.e.\ $400 \times
10^{-11}$. This would mean that the current model calculations need to
be way off, which seems not very likely in my opinion. We have learnt
a lot about the HLbL contribution from the theory side in the last 15
years and the model estimates did not change very much over time. In
fact the very recent preliminary and still incomplete evaluations
using DR's and lattice QCD seem to roughly confirm the estimates from
model calculations, but have the potential to much better control the
uncertainty.

% BibTeX or Biber users please use (the style is already called in the
% class, ensure that the "woc.bst" style is in your local directory) 
% \bibliography{name or your bibliography database}

\end{document}